# 5 Years of Defocused Observations of Exoplanet Transits with T100: Timing Perspective

**Baştürk, Özgür[1]; Yalçunkaya, S.[1]; Keten, B.[1]**

1Ankara Universitesi Fen Fakultesi Astronomi Bol. E Blok, 205, Tandogan Ankara, TR-06100 Turkey, obasturk@ankara.edu.tr, selcuk_yalcinkaya@yahoo.com, burakketen91@gmail.com

**Abstract**

We have been carrying out a program for over five years to observe transits of selected exoplanets with 1 meter Turkish Telescope, T100 (Baştürk et al. 2014, 2015), by making use of the well-established defocusing technique (Southworth et al. 2009) to achieve high photometric precision. In this contribution, we review the results of our observing program in timing perspective. The basic idea behind defocusing technique is to have the advantage of posing the detector for longer durations in the observations of bright stars, otherwise observed within very short integration times. Then the effect of the photon noise is diminished, which dominates in the short cadence observations. Longer exposures also help in reducing the noise contribution of the atmospheric scintillation. Noise contributions of the imperfect tracking and flat-fielding are mitigated by integrating over a larger area on the detector as well. Although fewer images can be acquired in a given time, we argue that the timing precision is improved because of better photometric precision. This contribution has been supported by TÜBİTAK-3001 project 116F350.

*Keywords: exoplanets, transit, TTV, photometry, telescope defocusing*

## INTRODUCTION

Precise measurements of the timings in an exoplanet transit light curve is crucial for several reasons. Mid-transit time is one such timing which should be determined with high precision in order to tell whether it increases at a constant rate with the orbital period or the rate of change is variable for some reason. In the presence of an additional but an unseen planet in the system, timing of the mid-point in a transiting exoplanet's light curve will fluctuate because of the gravitational perturbation of the additional third body (Ballard et al. 2011), and to a greater extent, the light time effect (LiTE) due to the orbital motion of the observed system around the center of mass with the unseen one (Holman & Murray 2005). This makes the detection of the unseen body possible from the Transit Timing Variations (TTVs) of the observed transiting exoplanet. In fact, several non-transiting planets have been detected so far thanks to TTVs they caused (Dawson et al. 2012, Saad-Olivera et al. 2017 and references therein) and the validity of the technique have been confirmed for multi-body systems with more than one transiting bodies (for Kepler-47 system see Vanderburg et al. 2017). These observations help theorists in their work toward understanding the formation of planetary systems. Tidal interaction between a close-in planet and its host star also causes an orbital period change, which is secular naturally, due to angular momentum considerations (Murgas et al. 2014).

Other important timings, that should be determined very precisely are the timings of the contact points. In the presence of a difference between the total egress and total ingress durations during an exoplanet transit, one should suspect of a difference in the speed of the orbital motion, hence an eccentric orbit. More than half of the exoplanets discovered from their transit signal are hot-Jupiters orbiting very close to their host stars on eccentric orbits, which was not expected at the beginning and posed a serious challenge for the theorists (Dawson et al. 2014).

Most ground-based transit surveys look for planets around bright stars (m < 13m) due to the limits posed by the instrumentation and the Earth's atmosphere. The objects of Kepler space mission (and now K2 mission), on the other hand, are faint stars in a certain region in the sky, selected to avoid the Sun entering the Field of View and more importantly, to increase the detection probability by increasing the number of solar like stars in the field. This helped in detecting more than 3000 planet candidates since 2009 but made it difficult to confirm the planetary status by complementary radial velocity observations. However, some of the next generation space surveys (TESS, Cheops, and PLATO) are going to search for planets around bright objects to be able to increase the SNR required in the spectroscopic studies of their atmospheres with the upcoming space missions like James Webb Space Telescope (JWST). These space-based surveys as well as other space telescopes such as GAIA will only have a very limited time to observe the transits of the detected planets. Hence the follow-up observations from the ground will have utmost importance for both characterization of these planets & their orbits, and the detection of the additional unseen bodies gravitationally bound to the systems from the variations in their transit timings (TTVs).

The number of planet candidates detected from their transit signals have already exceeded 4000, including those detected by the Kepler and K2 missions (http://exoplanet.eu). With the addition of aforementioned surveys, these numbers will increase exponentially. Hence, the ground-based efforts of small university observatories, and even the amateur observers will be required because they will be able to provide data on a much longer time basis, which will be crucial in the detection of TTVs induced by additional planets. This will constitute an important data source for theorists working on multi-planet system architectures, planet formation and migration mechanisms.

**TELESCOPE DEFOCUSING TECHNIQUE**
Since TTV amplitudes can be on the order of a few tens of seconds and contacts are very short-duration events, observers aim at precise as well as accurate measurement of the timing of each and every point in a transit light curve. Even the stratum-2 level GPS coordinated timing accuracy is satisfactory for such kind of work, because it provides sub-second timing precision. This precision can be achieved with very modest, inexpensive GPS devices. In ground-based observations, the quality of photometry is the major source of the scatter in transit light curves and hence dominates the error budget in the timing measurements. Therefore an observer should also aim at improving the quality of the photometry (Baştürk et al. 2014).

Using low-grade CCDs with high quantum efficiency over a wide range of wavelengths and larger apertures are obvious but expensive solutions to increase the Signal-to-Noise Ratio (hereafter SNR) in photometric measurements. These high-cost instruments are not in the reach of most small university observatories and advanced amateur astronomers, who can contribute significantly to the field. So intelligent observation techniques should be employed to get the most out of already existing observing equipment.

Telescope defocusing is a well-established technique that improves the photometric quality by decreasing the effect introduced by the photon noise and imperfect flat fielding (Southworth et al. 2009). Transits of the planets orbiting bright stars are performed at high cadence even with modest aperture size telescopes. Photon noise then becomes the major noise contributor due to the short exposure times employed to avoid the non-linear response of the CCDs or the saturation of the pixels in the worst case scenarios. In order to increase the exposure durations, the distance between the focal plane of the telescope and the detector can be changed to increase the number of pixels covered by the Point Spread Function (PSF) of the target, and hence decrease the number of photons hitting a given pixel in a given time. This also mitigates the noise introduced by the imperfect flat fielding in the order of a magnitude just because the significant inter-pixel response differences are being averaged out when a larger area on the detector is at hand. Anything that changes the position of the stars on the CCD (atmospheric scintillation, tracking errors, focus changes, seeing variations etc.) have lesser effects for the same reason. The effect of atmospheric turbulence can be very dramatic in the short run, whereas it is time-averaged during long exposures. Another advantage of posing the detector for longer durations is that the much longer fraction of an observing run is spent on collecting photons rather than reading-out the recorded images. Readout time becomes a real issue when bright stars and larger CCDs are considered. For the 4096x4096 CCD attached on the 1 meter Turkish Telescope, T100, located in TÜBİTAK National Observatory of Turkey, the readout time in unbinned mode is 45 seconds. In the observation of a $10^m$ star in a dark night with T100, the CCD should be exposed less than 5 seconds in the focused mode to avoid saturation let alone the non-linear response limit. This causes a very good fraction of the total observation time to be lost due to readout, hence called the "dead time".

The immediate drawbacks are poorer time resolution, increased noise contribution of the sky brightness and the readout procedure due to integration over a larger number of pixels. Sky brightness becomes the dominant term after some point, hence it constitutes the limit of both the exposure time and the size of the defocused image in the absence of intruding near-by objects to the targets. The noise contribution of the sky ($N_{sky}$) as given by Southworth et al. (2009) is given with the Equation 1, where texp is the exposure time, $n_{pix}$ is the number of pixels, $C_{sky}$ is the total count from the sky in ADUs. Hence the upper limit of the sky noise contribution depends heavily on the exposure time as well as the number of pixels in the aperture.

$$N_{sky} = \sqrt{t_{exp}.n_{pix}C_{sky}} \quad (1)$$

Until this limit, SNR value increases since the signal is increased linearly by telescope defocusing while the background noise ascends with the square root of the number of photons. Although the noise contribution of the readout procedure fluctuates throughout an observing session, the amplitude of the variation around the mean is negligible. Hence correcting with a constant readout noise value is sufficient in most of the cases. Poorer timing resolution can be healed by reading from a smaller area on the detector where the target stars are located. Such an option is provided by most of the image acquiring softwares used in the observations. This limits the number of good comparison stars with similar brightness and color to the target star; however, when there are sufficient number of such stars in a smaller area on the detector the readout time significantly decreases.

### DEFOCUSED T100 OBSERVATIONS OF SELECTED TRANSITS

Full papers, including main body, figures and tables, should not be more than 8 pages. We have been carrying out defocused photometric transit observations of the selected exoplanets with the 1 meter Turkish telescope, T100, for over 5 years now. T100 is located in Bakırlıtepe / Antalya campus of the TÜBİTAK National Observatory of Turkey (TUG). We achieved sub-milimagnitude level precision in several of our transit observations for the brightest host stars in our sample (Baştürk et al. 2014, 2015). The telescope has a quality CCD (SI1100 Cryo model with a Fairchild 486 BI chip) with 4096 by 4096 pixels and a relatively wide Field of View (20 x 20 arcminute-squared) making it possible to include sufficient number of reference stars for ensemble photometry (Honeycutt 1992). We observed our targets in Cousins R band (Rc) in most of our observations, and also made use of Cousins I (Ic), and Sloan r, i, and z bands for a few cases because we observed planet-hosting cool dwarf stars. In addition, our observations are less affected by the limb darkening, which is weaker at longer wavelengths.

In the determination of the exposure times, we take the total duration of the transit into account. We aim at acquiring at least 50 images per transit, changing the focus to decrease the exposure time when necessary especially during the ingress and egress times. We heavily defocus the telescope in longer duration transits to the limit the sky noise starting to dominate in the absence of close-by sources, which can further limit the extent of the defocusing to avoid broadened PSF profiles of two close sources mix with each other. The maximum exposure time within these limits has been achieved in the transit observations of XO-3b, whose host star is $9^m.86$ and transit duration is 10380 seconds. The readout time for the 16 megapixel CCD is 45 seconds, which allowed us to obtain 53 in-transit images (79 in total) for this particular transit (Figure-1). We pay attention to keep the counts in the linear regime of our CCD (25000 – 35000 ADU count-range in our case). We also pay attention to keeping the size of the defocused image at a certain value (~ 150 pixels in diameter for XO-3b transits) to integrate the counts in a fixed aperture size for all the images during aperture photometry.

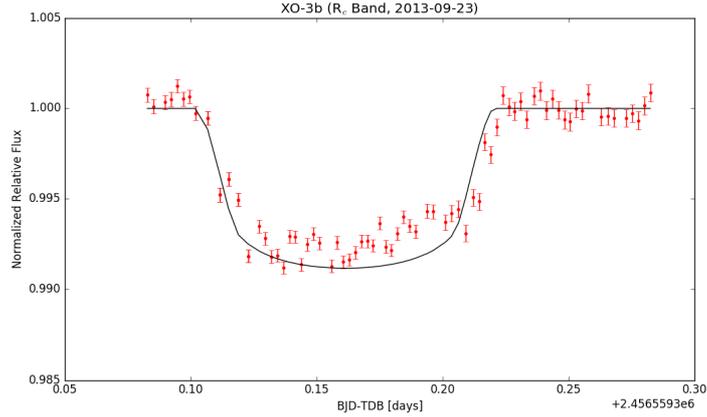

**Figure 1.** Transit light curve (rms = 0.0010 in normalized units) of XO-3b, observed with T100 on 23 September 2013 in $R_c$ filter. The transit model (in black) is fitted by fixing the planetary and stellar parameters to that from Winn et al. (2008) and adjusting only the mid-transit time.

DATA ANALYSIS AND TRANSIT LIGHT CURVE FITTING

We analyze our data in a homogeneous way using AstroImageJ (reference) for both reductions of our CCD images and ensemble differential photometry. We select the comparison stars with similar brightness to the target in the given passband to form the synthetic comparison from the ensemble. We weight the individual comparison stars by the scatter in their light curves. We perform differential aperture photometry in the usual manner (reading sky counts from an annulus, assigning a certain value to sky brightness, and having an instrumental magnitude for the target), only using a synthetic comparison star rather than a real star. We then detrend the light curves from the airmass effect and normalize them to the out-of-transit light level.

In the transit light curve modeling, we adopt the essential stellar spectroscopic parameters (effective temperature, $T_{eff}$; surface gravity, log g; and metallicity, [Fe / H]) as well as the orbital parameters (eccentricity, e; argument of periastron, ω) from the most thorough spectroscopic study, which is the discovery paper in most cases. We use quadratic limb darkening coefficients from Claret & Bloemen (2011)'s tables and interpolate for the stellar parameters (Eastman et al. 2013) with the help of a web-based applet[1]. In the absence of precise and accurate values of the transit-related parameters (the ratio of the radii of the planets-to-stars, $R_p / R_s$; semi-major axes scaled to the stellar radii, $a / R^*$; and the orbital inclinations, i), we make use of the simple modeling tool provided by the AstroImageJ software. If these parameters have already been calculated from the light curves more precise than our own, we fix their value in the modeling as well. We select Barycentric Dynamical Julian Date (BJD-TDB) as our reference time scale in the determination of mid-transit times and measure the mid-points with the method employed by AstroImageJ software (Collins et al. 2017).

---
[1] http://astroutils.astronomy.ohio-state.edu/exofast/limbdark.shtml

## RESULTS

We obtained precise transit observations of selected exoplanets (WASP-103b, WASP-37b, HAT-P-37b, WASP-43b, HAT-P-23b, Qatar-1b) in our sample with 1 meter Turkish telescope T100, making use of the defocusing technique. We reduced and analyzed our precise light curves as described in the previous section. We present our light curves in Figure 2 to Figure 7 and summarize our results in Table-1.

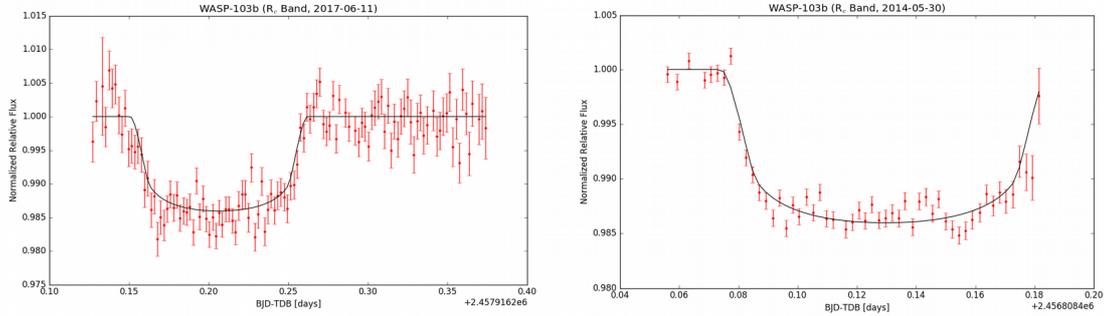

Figure 2. Transit light curves of WASP-103b on 11 June 2017 and 30 May 2014 in $R_c$ filter. The transit model (in black) is fitted by fixing the planetary and stellar parameters to that from Gillon et al. (2014) and adjusting only the mid-transit time.

Table 1. Information on T100 Defocused Observations. WASP-37b, HAT-P-37b, HAT-P-23b, and Qatar-1b had to observed in 2x2 binning to avoid dead time due to readout. Radii of the defocused star images are given in pixels. RMS values have been computed in the usual manner with respect to the light curve models. Errors in mid-times have been converted into seconds for better comprehension.

| Planet | V Mag. | Obs. Date | Exp. Time | Defocus Radius | RMS | Tc (BJD-TDB) | Error in Tc |
|---|---|---|---|---|---|---|---|
| WASP-103b | $12^m.1$ | 11.06.2017 | 120 s | 20 pixels | 0.0027 | 2457916.406243 | 72 s |
| WASP-103b | $12^m.1$ | 30.05.2014 | 135 s | 25 pixels | 0.0014 | 2456808.529636 | 66 s |
| WASP-37b | $12^m.7$ | 27.04.2017 | 135 s | 15 pixels | 0.0024 | 2457871.474758 | 60 s |
| HAT-P-37b | $13^m.2$ | 04.08.2015 | 150 s | 30 pixels | 0.0022 | 2457239.481626 | 40 s |
| WASP-43b | $12^m.4$ | 14.04.2015 | 60 s | 25 pixels | 0.0023 | 2457127.345756 | 22 s |
| HAT-P-23b | $12^m.0$ | 25.09.2014 | 135 s | 35 pixels | 0.0013 | 2456926.300776 | 35 s |
| Qatar-1b | $12^m.8$ | 17.08.2014 | 120 s | 30 pixels | 0.0011 | 2456887.314251 | 20 s |

## CONCLUSION AND FUTURE PROJECTION

Our results summarized in Table 1 prove strength of telescope defocusing technique in terms of photometric precision, which is reflected as in timing precision. The errors of the mid-transit times, derived from our defocused T100 observations are less than a minute in 5 of 7 observations, and only 12 seconds more than a minute in the worst case (for WASP-103b on 11.06.2017 observation). This precision level is quite satisfactory for Transit Timing Variation (TTV) studies to be able to detect small-size planets on relatively larger orbits (Dawson et al. 2012, Saad-Olivera et al. 2017, Vanderburg et al. 2017). Contributing to the discovery of a potential exoplanet can motivate small university observatories, astronomy students, and advanced amateur astronomers, who can reach small-to-medium size telescopes with relatively modest equipment.

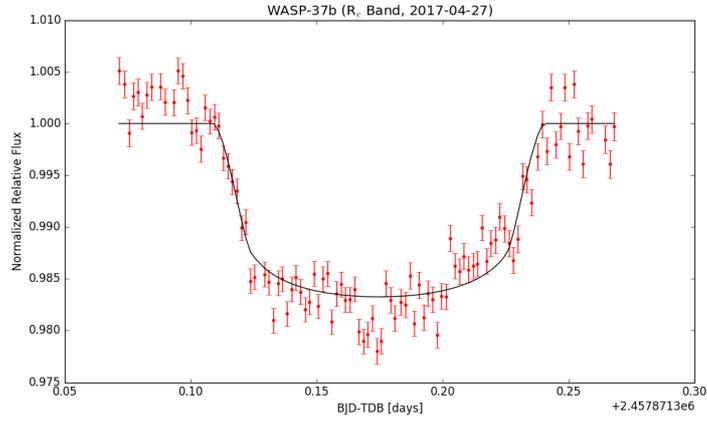

**Figure 3.** Transit light curve of WASP-37b on 27 April 2017 in Rc filter. The transit model (in black) is fitted by fixing the planetary and stellar parameters to that from Simpson et al. (2011) and adjusting only the mid-transit time.

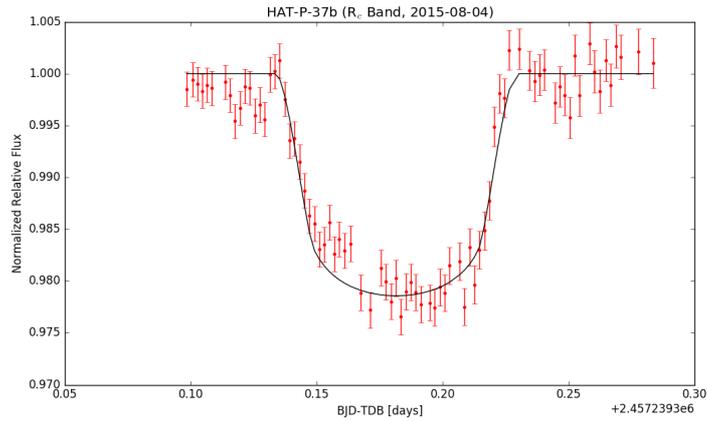

**Figure 4.** Transit light curve of HAT-P-37b on 4 August 2015 in Rc filter. The transit model (in black) is fitted by fixing the planetary and stellar parameters to that from Bakos et al. (2012) and adjusting only the mid-transit time.

From this perspective, we started a project to search for TTVs in systems with at least one transiting planets. We have been observing transits of exoplanets, orbiting on eccentric orbits and displaying radial velocity residuals from their best fits with the 1 meter Turkish telescope T100. We achieve high photometric precision, which leads to high precision in mid-transit timing measurements with the well-established telescope defocusing technique even in the observations of relatively faint objects of ground-based transit surveys.

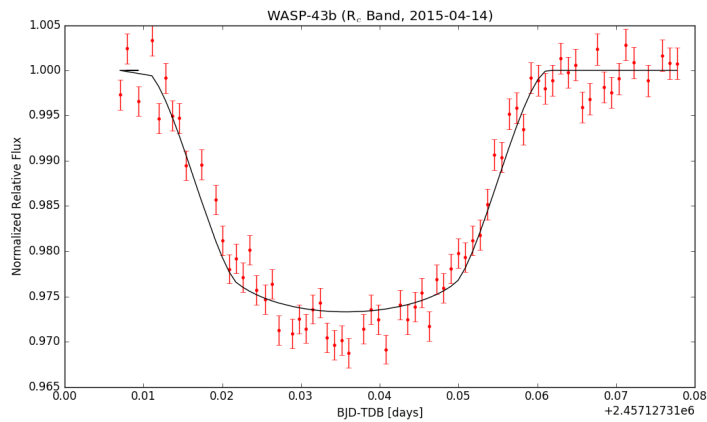

**Figure 5.** Transit light curveof WASP-43b on 14 April 2015 in Rc filter. The transit model (in black) is fitted by fixing the planetary and stellar parameters to that from Gillon et al. (2012) and adjusting only the mid-transit time.

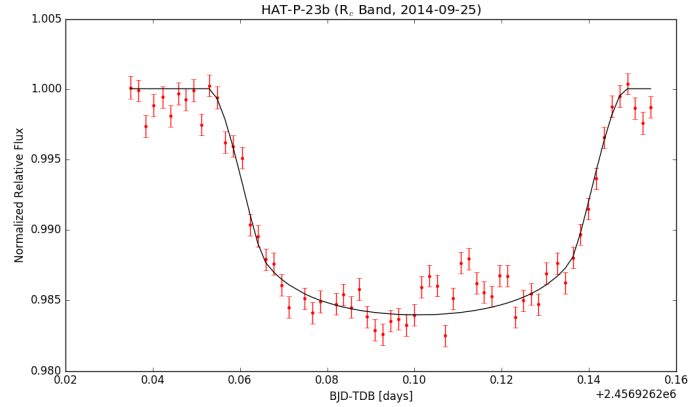

**Figure 6. Transit light curve of HAT-P-23b on 25 September 2014 in Rc filter. The transit model (in black) is fitted by fixing the planetary and stellar parameters to that from Bakos et al. (2010) and adjusting only the mid-transit time.**

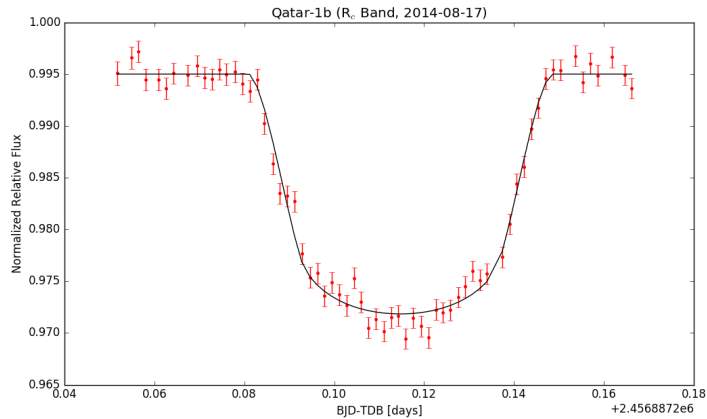

**Figure 7. Transit light curve of Qatar-1b on 17 August 2014 in $R_c$ filter. The transit model (in black) is fitted by fixing the planetary and stellar parameters to that from Alsubai et al. (2011) and adjusting the mid-transit time as well as the orbital inclination.**

In addition our own observations, we started collecting light curves from the literature, and from databases like Extrasolar Planet Database (ETD in short, http://var2.astro.cz/ETD/index.php) for our targets. After measuring mid-transit times in a homogeneous way with the same technique in the same time reference, we are refining the transit ephemerides (orbital period and the reference epoch) for our targets from least squares linear fits to the mid-transit timing measurements weighted by their timing uncertainty. We then plot the change in mid-transit time determined with respect to an arbitrarily chosen one against time (epoch) and visually inspect the cases with potential TTVs. We are also performing frequency analyses on the data to understand if there is a dominant frequency. In the cases, where a significant peak (over 4 σ) is detected in the frequency analysis, we compute the phases with respect to the period corresponding to the detected dominant frequency. You can see two such individual cases (XO-3b and HAT-P-19) in two separate poster presentations (Yalçınkaya et al. 2018, Keten et al. 2018) in this workshop presented by the master's students in our research group. Our results showed that telescope defocusing brings a significant advantage over the traditional observing method, especially when the errors of the transit mid-times are considered.

ACKNOWLEDGMENT

This contribution has been supported by TÜBİTAK-3001 project 116F350. We also acknowledge TÜBİTAK for a partial support in using T100 telescope with the project numbers 12CT100-378, 16AT100-997, 16BT100-1034, 16CT100-1096, 17BT100-1196.